# Fundamental constraints for the length of the MOSFET conduction channel based on the realistic form of the potential barrier

**Maksym V. Strikha[1, 2], Anatolii I. Kurchak[2], and Anna N. Morozovska[3],**

[1] Taras Shevchenko Kyiv National University, Faculty of Radiophysics, Electronics and Computer Systems, pr. Akademika Hlushkova 4g, 03022 Kyiv, Ukraine

[2] V. Lashkariov Institute of Semiconductor Physics, National Academy of Sciences of Ukraine, pr. Nauky 41, 03028 Kyiv, Ukraine

[3] Institute of Physics, National Academy of Sciences of Ukraine, pr. Nauky 46, 03028 Kyiv, Ukraine

The work estimates the minimum channel length of the MOSFET transistor, which is the bacis device of modern electronics. Taking into account the real shape of potential barrier in the channel shows that the electron tunnels through a region significantly shorter than the physical length of the channel $L$ in the presence of drain voltage, and so the available estimate of the minimum quantum constraint channel length in silicon MOSFET, $L_{min} \approx 1.2$ nm, is significantly underestimated. The fact makes it clear why after reaching 5 nm working lengths of the channel it was impossible to reach the long-declared values of 3 nm under maintaining the proper level of functionality of the transistor. The estimates made in this work confirm that the fundamental scaling limits of silicon MOSFETs have almost been reached.

Corresponding author: maksym.strikha@gmail.com

Metal-Oxide-Semiconductor Field-Effect Transistor (MOSFET) is the basic device of modern electronics. Therefore, it is still the subject of numerous experimental and theoretical studies (see, for example, [1]). The physics of the MOSFET is determined by how the electrons move from the source "S" through the conductive channel to enter the drain "D". The electron current $I_D$ through the conduction channel of the transistor with the appropriate level of functionality is effectively controlled by gate "G", which is isolated from the conduction channel by a dielectric layer. This control is based on the fact that the gate, depending on the applied voltage, changes the height of the potential barrier between the source and the drain.

There is a permanent miniaturization (i.e. "scaling") of MOSFETs: when at the end of the 20th century the standard length of the channel between the source and drain was usually about 100 nm, today it is already about 10 nm. At the beginning of the 21st century, the creation of experimental samples with much shorter channels was announced: at first of the 6 nm length [2], and then of 3 nm length [3]. The thickness of the oxide layer in modern devices can already be less than 2 nm.

However, over the last two decades, it has become more and more clear that Moore's empirical law, which satisfactorily described the situation with transistor scaling since 1965, assuming doubling the number of elements on the chip every 24 months, is close to its exhaustion. It occurs due not only to the numerous technological problems of scaling (see, e.g. Ref. [4] and refs. therein), but also to the fundamental limitations associated with the quantum nature of electrons in nanosystems.

Back in 1961, Rolf Landauer theoretically predicted the presence of the minimum energy required to switch the system from "ON" to "OFF" modes. Although Landauer [5] proceeded from the consideration of the uncertainty principle, the same result can be achieved by simple visual considerations, based on the examination of the MOSFET band structure. **Figure 1a** shows the energy of the conduction band bottom as a function of the x coordinate along the channel of the *n*-channel MOSFET in the "ON" mode. The high positive voltage applied to the gate virtually destroys the potential barrier and the electrons from the source pass through the conduction channel to the drain. We assume that the electro-transport in the channel is ballistic, so the electrons give off their energy to the drain and relax due to intense inelastic interactions in the drain terminal.

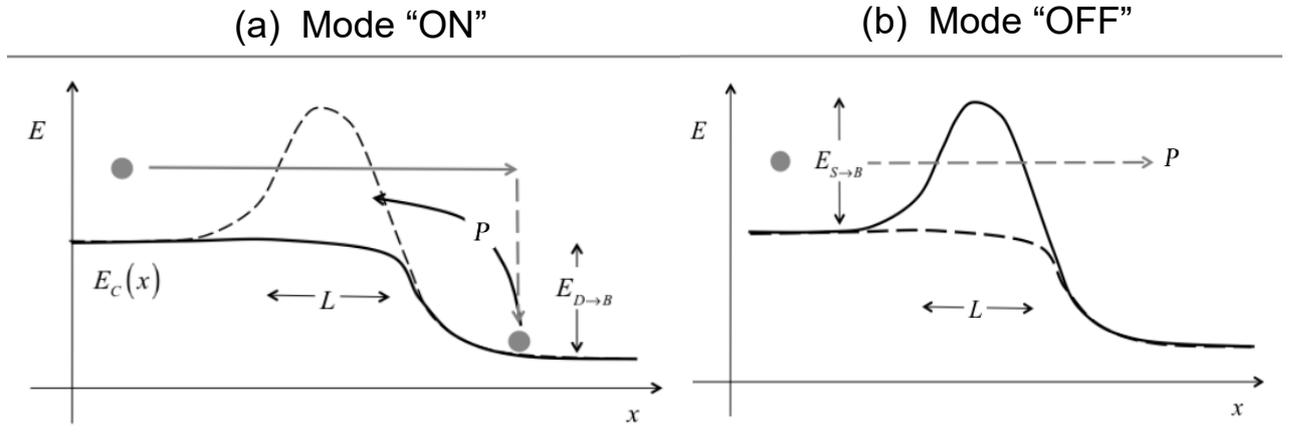

**Figure 1. (a)** Switching the MOSFET from one mode to another: a solid line corresponds to the mode "ON", dotted to the mode "OFF". The switching probability is $1 - P$, where $P$ is the probability of thermoelectronic remission from the drain to source. **(b)** In the "OFF" mode, there is a probability P of quantum mechanical tunneling of the electron through the barrier with a width $L$. Adapted from Ref.[1].

This simple model allows to estimate the minimal energy required to switch from one mode to another. High gate voltage in the "ON" mode eliminates the barrier between the source and the channel, but the barrier $E_{D \to B}$ between the drain and the top of the barrier remains, because a positive voltage is applied to the drain. Because the electrons thermally relax in the drain and transfer their kinetic energy to the lattice phonons, there is still some possibility $P$ that the electrons still overcome the barrier $E_{D \to B}$ and return to the source. In this case a switching will not occur. Requiring that this probability be less than $1/2$, we obtain

$$P = e^{-E_{D \to B}/kT} < \frac{1}{2}, \qquad (1a)$$

from here we estimate the minimal switching energy as:

$$E_{min} \equiv E_S\big|_{min} = kT \ln 2, \qquad (1b)$$

which is 0.017 eV at room temperature. These simple considerations are heuristic, but both Landauer's pioneering work [5] and modern detailed analysis [6] lead to the same results for the value of the minimal switching energy.

Within the framework of the same simple scheme [1] it is possible to estimate the minimal length of the MOSFET conduction channel. As is clear from **Fig. 1b**, the height of the barrier in the "OFF" mode must be at least not less than $E_{min}$, which serves as a guarantee that the electrons will overcome the barrier with a probability less than 1/2. In this case, the minimal width of the barrier (channel length) is determined by quantum mechanical tunneling through it. The probability that an electron from a source tunnels through a barrier can be estimated in the Wentzel-Kramers-Brillouin approximation (WKB, see e.g. [7]), which gives a known formula for the probability of tunneling of a particle with energy $E$ and mass $m^*$ through barrier with potential $V(x)$ between the points $x_1$ and $x_2$:

$$P \approx exp\left(-\frac{2}{\hbar}\int_{x_1}^{x_2}\sqrt{2m^*(V(x)-E)}dx\right) \qquad (2)$$

From the requirement that the tunneling probability should be less than ½ in the "OFF" mode, it follows that

$$P = exp\left(-\frac{2}{\hbar}\int_{x_1}^{x_2}\sqrt{2m^*E_{S\to B}}dx\right) < \frac{1}{2}. \qquad (3)$$

In Eq.(3), we considered the rectangular potential barrier along the entire length $L$ of the channel. By putting it equal to the minimal value $E_{min}$ described in (1b), we obtain the minimal length of the conduction channel:

$$L_{min} \approx \frac{ln2}{2}\frac{\hbar}{\sqrt{2m^*E_{min}}} \qquad (4)$$

Let us estimate the value of this minimal length for a thin inverse *n*-type channel in Si (100), where a quantization is already taking place in the direction into the substrate. As it can be shown [8], the lower subzone with $n=1$ corresponds to the effective mass $m_l^* = 0.92m_0$ in the direction of localization and valley degeneration 2. But the effective mass in the direction $x$ of free motion along the plane of the channel is $m_t^* = 0.19m_0$, and it must be taken into account in Eq.(4).

This leads to a value $L_{min} \approx 1.2$ nm that may give the impression that there is a possibility to reduce the channel lengths in silicon MOSFETs compared to those already used in electronics, at least a few more times. However, the real problems arising now under the scaling of transistors doubts this statement. After all, although the 3 nm length of experimental channels was reached more than 15 years ago [3], they have not yet become a practice in electronics. One reason may be that a simple estimate (4) is made without taking into account the realistic shape of potential barrier in the channel, in fact assuming that the drain voltage is absent.

However, this potential is no longer rectangular under the application of direct displacement voltage to the drain, but has a form similar to that shown in **Fig. 1b**. In this case, as it is well-known from the rigorously constructed theory of nanotransistors [1], a relatively narrow region near its apex with length $\ell \ll L$ is critical for the classical transmission of the barrier, where the potential at the source varies slightly and the electric field is almost absent [9]. An electron that has been able to pass through this narrow region is dragging out by a strong electric field to the drain, even participating in collisions. Therefore, the realistic potential corresponding to the situation of application of voltage V to the drain is approximated by the form shown in **Fig. 2,** and obeys the expression:

$$V(x) = \begin{cases} E_{min}, & 0 \leq x \leq \ell, \\ E_{min} - \frac{E_{min}+|eV|}{L-\ell}(x-\ell), & \ell \leq x \leq L. \end{cases} \qquad (5)$$

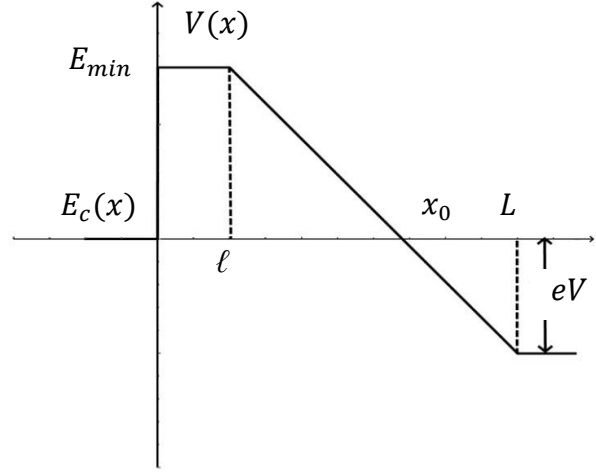

**Figure 2**. Approximation of the potential barrier in the MOSFET channel of length $L$. Electrons tunnel through the barrier from the source to drain (from left to right). Voltage $V$ is applied to the drain.

A similar approximation is widely used to consider the problems of carrier transport through the MOSFET with scattering [10]. It should be noted that the length $\ell$ is a function of the gate voltage $V$, and $\ell \to L$ at $V \to 0$ (it is clear that without voltage the whole channel becomes a region with a constant potential and zero electric field). However, as shown by numerical simulations [11], the voltage increase very quickly decreases $\ell$ to a certain constant value, and in the wide range of drain voltages we can assume that $\ell = \xi L$, where the parameter $\xi \approx 0.1$.

Further, it is necessary to integrate Eq.(3) taking into account the expression (5) for $V(x)$ from 0 to $x_o$, which corresponds to zero value of the second expression in Eq.(5), because the tunneling through the barrier shown in **Fig. 2** is possible only for positive electron energies. This integral is taken rigorously and leads to the modification of Eq.(4):

$$L_{min}(\chi) \approx \frac{1}{2} ln2 \frac{\hbar}{\sqrt{2m^*E_{min}}} \frac{1}{\chi}, \qquad (6)$$

Where $m^*$ is an electron effective mass, $E_{min} \cong kTln2$ is given by Eq (1b) and $\hbar$ is a Plank constant. The function:

$$\chi(U,\xi) = \xi + \frac{2}{3}\frac{1-\xi}{1+U}, \qquad (7)$$

where $U = |eV|/E_{min}$ is the dimensionless ratio of applied voltage to $E_{min}$.

**Figure 3a** shows the dependence of the parameter $\chi$ on the dimensionless voltage ratio $U$ for several different values of $\xi$, which order of magnitude corresponds to the value obtained by numerical simulations [11]. As it can be seen, we obtain the value $\chi \sim 0.3$, which relatively weakly depends on the further voltage increase in the range of voltages for which the expression (7) is valid (recall that for very small voltages the parameter $\xi$ is also a function of voltage and trends to 1 for V $\to 0$). Taking into account the functional form (6), this triples the above value of the minimal channel length $L_{min} \approx 1.2$ nm.

It should be noted that numerical simulations [11] were performed for sufficiently long channels with $L > 30$ nm. It can be assumed that for ultrashort channels with L about several nanometers, the value of the parameter ξ may be slightly larger. The results of χ calculations for in this case are shown in **Fig. 3b.**

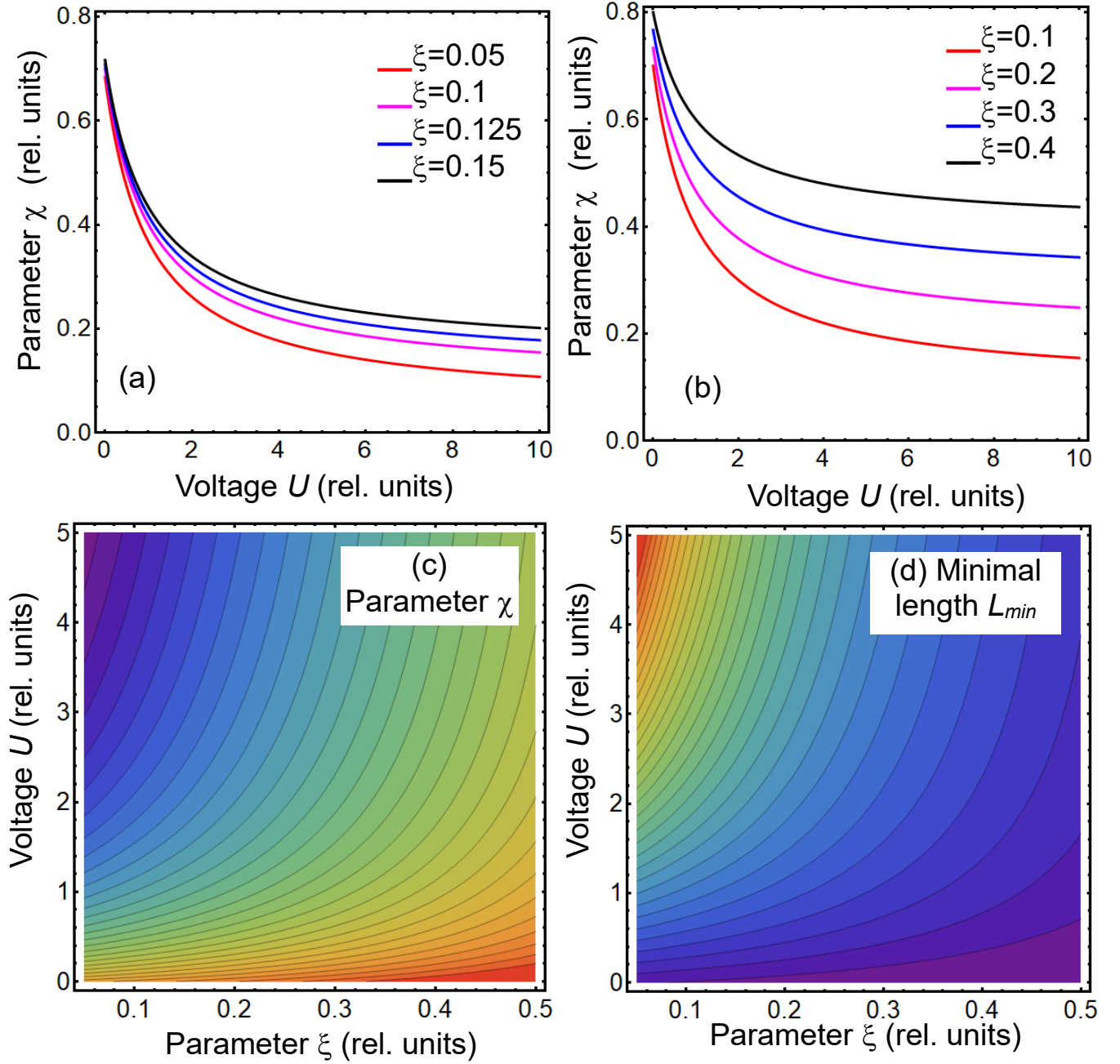

**Figure 3.** (a) Dependence of the parameter χ on the ratio $U = |eV|/E_{min}$ for small ξ values. (b) Dependence of the parameter χ on $U$ for different hypothetical ξ-values, which may correspond to the situation in ultrashort channels. (c) Contour map of parameter $\chi(U, \xi)$, which scales from 0.1 (violet color) to 0.8 (red color). (d) Contour map of $L_{min}(U, \xi)$, which scales from $1.2L_0$ (violet color) to $6.43L_0$ (red color).

As can be seen from **Fig. 3b**, even the knowingly overestimated value of ξ = 0.4 doubles the estimate $L_{min} \approx 1.2$ nm, which allows us to understand why the transistors with a 3-nanometer channel [3] were ultimately nonfunctional.

Contour map of the parameter $\chi(U, \xi)$ and minimal channel length $L_{min}(U, \xi)$ in coordinates $(U, \xi)$ are shown in **Fig. 3c** and **3d**, respectively. The parameter $\chi(U, \xi)$ scales from 0.1 (violet color) to 0.8 (red color), and the inverse function, $L_{min}(U, \xi)$, scales from $1.2L_0$ (violet color) to $6.43L_0$ (red color), where $L_0 = \frac{1}{2} \ln 2 \frac{\hbar}{\sqrt{2m^* E_{min}}}$.

These results correlate with the conclusions of quantum transport modeling in Si NW MOSFET [12]. In this work, it was shown that at the length of the gate part of the channel, $L_G \approx 12$ nm, almost all the current $I_{OFF}$ is flowing higher than the barrier. Such a transistor operates in the usual classical mode, controlled by a barrier. As the length of the channel decreases to 10 nm, a small number of electrons already tunnel through the barrier, but the normal classical mode controlled by the barrier still dominates. At $L_G \approx 7$ nm a significant part of the electrons that cause the current $I_{OFF}$ is already tunneling through the barrier. Finally, at $L_G \approx 5$ nm, the significant part of the current $I_{OFF}$ is caused by tunneling through the barrier. It is impossible to control the current by controlling the barrier height at such small lengths of the conduction channel, because the barrier has become permeable to electrons.

The results obtained in this work are in agreement with the generally accepted today idea that the classical model of MOSFET transport is quite applicable to Si transistors with a conduction channel length of up to 10 nm and even slightly less. However, further scaling to 5 nm already poses serious problems, both technological ones related with the increasing the role of parasitic resistors and capacitors for very short conduction channels, and fundamental ones related with the tunneling through the barrier. Numerical simulation of transistors with stressed substrates specifically oriented relative to the direction of the conduction channel, which allows to increase the of the effective mass value in Eq.(6), shows that it is likely possible to implement an acceptable mode of MOSFET operation with conduction channel length less than 5 nm [13]. However, this value will not be significantly reduced in the future due to the fundamental limitations associated with the quantum nature of the electron's motion through such a short channel.

**To conclude**, we estimate the minimal length of the channel of MOSFET, being the principal device of modern electronics. An account of the real form of the potential form in the channel demonstrates, that when voltage is applied to the drain, an electron is tunneling through the region, which is essentially shorter than the channel physical length *L,* and therefore the estimation for the minimal channel length in the Si MOSFET, present in literature, $L_{min} \approx 1.2$ nm, is essentially lowered. This explains, why, after reaching the working length of the channel of 5 nm the announced long time ago values of 3 nm had not been achieved yet with a proper level of the transistor operation functionality. The estimations of our work confirm that fundamental limits of Si MOSFET scaling have been almost reached already.

**Authors contribution.** M.V.S. generate the research idea, performed analytical calculations and wrote the manuscript. A.I.K. and A.N.M. prepared graphics.

**Acknowledgements.** A.N.M. acknowledges the National Research Foundation of Ukraine (Project number 2020.02/0027).